\documentclass[twocolumn,showpacs,amsmath,amssymb,prb,aps]{revtex4}

\usepackage{graphicx}
\usepackage{dcolumn}
\usepackage{bm}
\usepackage{amsmath}
\usepackage{amsfonts}
\usepackage{amssymb}
\usepackage{MnSymbol}
\usepackage{float}

\begin{document}

\preprint{JAP/NonlinearTiNResonators}

\title{Operation of a titanium nitride superconducting microresonator detector in the nonlinear regime}

\author{L.~J. Swenson$^{1,2,}$\footnote{Electronic mail : swenson@astro.caltech.edu}}
\author{P.~K. Day$^2$}
\author{B.~H. Eom$^{1}$}
\author{H.~G. Leduc$^2$}
\author{N. Llombart$^3$}
\author{C.~M. McKenney$^1$}
\author{O. Noroozian$^4$}
\author{J. Zmuidzinas$^{1,2}$}
\affiliation{$^1$Division of Physics, Mathematics, and Astronomy, California Institute of Technology, Pasadena,
CA 91125}
\affiliation{$^2$Jet Propulsion Laboratory, California Institute of Technology, Pasadena, CA 91109}
\affiliation{$^3$Delft University of Technology, 2628 CD Delft, The Netherlands}
\affiliation{$^4$Quantum Sensors Group, National Institute of Standards and Technology, Boulder, CO 80305}
\date{\today}

\begin{abstract}

If driven sufficiently strongly, superconducting microresonators
exhibit nonlinear behavior including response bifurcation.
This behavior can arise from a variety of physical mechanisms including heating effects,
grain boundaries or weak links, vortex penetration, or through the
intrinsic nonlinearity of the kinetic inductance.
Although microresonators used for photon detection are usually driven fairly hard in order
to optimize their sensitivity, most experiments to date have not explored
detector performance beyond the onset of bifurcation.
Here we present measurements of a lumped-element superconducting microresonator
designed for use as a far-infrared detector and operated deep into the nonlinear regime.
The 1~GHz resonator was fabricated from a 22 nm thick titanium nitride film with a critical temperature of
2 K and a normal-state resistivity of $100\, \mu \Omega\,$cm.
We measured the response of the device when illuminated with 6.4 pW optical loading
using microwave readout powers that ranged from the low-power, linear regime to 18 dB beyond the onset of bifurcation.
Over this entire range, the nonlinear behavior is well described by a nonlinear kinetic inductance.  The best noise-equivalent power of $2 \times 10^{-16}$ W/Hz$^{1/2}$ at 10 Hz was measured at the highest readout power, and represents
a $\sim$10 fold improvement compared with operating below the onset of bifurcation.

\end{abstract}


\maketitle

\section{INTRODUCTION}

\begin{figure}[ht]
   \centering
    \includegraphics[width=82mm]{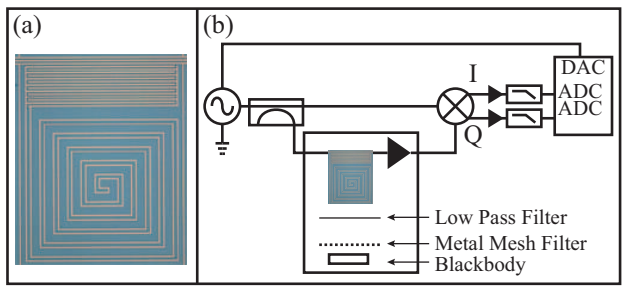}
      \caption{(Color online) Measurement setup. (a) Micrograph of a single
      LEKID-type microresonator detector from a 16x16 array.
      The interdigitated capacitor is visible above the spiral inductor/absorber.
      The array was fabricated from a TiN film with a critical temperature of $T_c = $2 K deposited on a high-resistivity silicon substrate.  The 22 nm film thickness was significantly thinner than the
      effective penetration depth
       $\lambda = \sqrt{\hbar \rho_n/(\mu_0 \pi \Delta_0) } \approx$ 750 nm.  (b) Electrical and optical setup.  The array is refrigerated to 110 mK and read out using a voltage-controlled microwave signal generator and standard homodyne detection electronics.  A
       variable-temperature
       blackbody is located
       behind a 4.2~K band-defining 215 $\mu$m metal-mesh filter (BW = 43 $\mu$m).  A 7~mm diameter aperture on the still shield is further equipped with 1.6 mm of high-density polyethylene and a 300 cm$^{-1}$ low-pass filter.  Various electrical signal-conditioning amplifiers and attenuators, including two 20 dB attenuators located on the 4.2 K and still stages, are not shown.
              }
         \label{figCircuitPixel}
\end{figure}

Superconducting detector arrays have a wide range of applications
in physics and astrophysics.\cite{irwin:63,Zmuidzinas:2004}
Although the development of multiplexed readouts has allowed array sizes to
grow rapidly over the past decade,
there is a strong demand for even larger arrays.
For example, current astronomical submillimeter cameras feature arrays with
up to $10^4$ pixels.\cite{hilton:513}
In comparison, the proposed CCAT 25-meter submillimeter telescope\cite{Astro2010}
would require more than $10^6$
pixels in order to fully sample the focal plane
at a wavelength of $\lambda=350~\mu$m.
Another example is cryogenic dark matter searches which currently implement
arrays with $<$50 pixels.
Each pixel features an integrated transition-edge sensor (TES) to detect athermal phonons
for $\sim$100-500~g of cryogenic detection mass.\cite{akerib:818, armengaud:329}
Proposed ton-scale searches would require of order $10^4$ detectors.
In order to meet these challenging scaling requirements,
it is highly desirable to simplify detector fabrication and to increase multiplexing factors
in order to reduce system cost.
From this perspective, superconducting microresonator detectors
\cite{Day:2003, Mazin:2004,mazin:2009,baselmans:292,Zmuidzinas:2012, Noroozian:2012}
are particularly attractive.
In these devices, the energy to be detected is coupled to a superconducting film,
causing Cooper pairs to be broken into individual electrons or quasiparticles,
which leads to a perturbation of the complex
ac conductivity
$\delta \sigma(\omega) = \delta \sigma_1 - j \delta \sigma_2$.
Very sensitive measurements of $\delta \sigma(\omega)$ may be made
if the film is patterned to form a microwave resonant circuit.
Because both the amplitude and phase of the complex transmission of the circuit
can be measured (see Fig.\ \ref{figCircuitPixel}), information on both
the dissipative ($\delta \sigma_1$) and reactive ($\delta \sigma_2$) perturbations
may be obtained simultaneously,
giving the user a choice of using reactive readout, dissipation readout, or both.
Frequency multiplexing of a detector array is readily accomplished by
designing each microresonator to have a different resonant frequency
and coupling all of the detectors to a single transmission line for excitation and readout.

Although various options exist for coupling pair-breaking photons or phonons into the superconductor,
the simplest approach for a number of applications is to directly illuminate the microresonator.
For good performance, the resonator must be designed to be an efficient absorber.
A particularly notable example
is the structure introduced by Doyle \textit{et al}
for far-infrared detection,
known as the lumped-element kinetic inductance detector or LEKID.\cite{Doyle:2007,Doyle:2008}
Fig.\ \ref{figCircuitPixel}(a) shows a variant of this concept,
designed for low interpixel crosstalk and polarization-insensitive operation.\cite{noroozian:1235}
The structure consists of a coplanar stripline spiral inductor and an interdigitated capacitor.
The microwave current density in the inductor is considerably larger than in the capacitor,
so the inductor is the photosensitive portion of the device.
The resonant frequency can be tuned simply by varying the geometry of
the capacitor or inductor during the array design.
Further, only a single lithography step is necessary for pattering an array of resonators
from a superconducting thin-film deposited on an insulating substrate.
The simplicity of these devices
has led to the demonstration of prototype arrays suitable
for submillimeter astronomy.\cite{monfardini:24}
Similar devices have been developed for optical astronomy
\cite{Mazin:2010} and dark matter detection experiments.\cite{swenson:263511, moore:232601}

Because the pixel size is comparable to the far-infrared wavelength, the details of
the resonator geometry do not strongly affect the absorption of radiation.
However, in order to achieve high absorption efficiency, the
effective far-infrared surface resistance of the structure should be around
$R_\mathrm{eff} = 377\,\Omega/(1 + \sqrt{\epsilon_r}) \approx 86\,\Omega$
when using a silicon substrate with dielectric constant $\epsilon_r \approx 11.5$.
This results in an approximate constraint on the sheet resistance $R_s$ and area filling factor $\eta_A$
of the superconducting film, $R_\mathrm{eff} \approx R_s / \eta_A$,
which is straightforward to satisfy if a high-resistivity superconductor such as TiN
is used.\cite{Leduc:2010}
These considerations provide a starting point for pixel design;
detailed electromagnetic simulations may then be used to optimize the absorption.
TiN is a particularly suitable material for resonator detectors due to its high intrinsic quality $Q_i$ which can exceed $10^6$ and a tunable $T_c$ based on nitrogen content ($0 < T_c < 4.7$\ K).

In practice, superconducting microresonators exhibit excess frequency
noise.\cite{Day:2003,Mazin:2004,Gao:2008,Zmuidzinas:2012}
This noise is due to capacitance fluctuations\cite{Noroozian:2009}
caused by two-level tunneling systems that are known to be present in
amorphous dielectrics.\cite{Anderson:1972,Phillips:1972}
Such material is clearly present when deposited dielectric films are used in the
resonator capacitor.\cite{Martinis:2005}
However, experiments have shown that even when
the capacitor consists of a patterned superconducting film on a high-quality
crystalline dielectric substrate, a thin surface layer of amorphous dielectric material is still present
and causes excess dissipation and noise.\cite{Gao:2008a,Gao:2008b}
This two-level system (TLS) noise has been studied extensively and a number of
techniques have been developed to reduce it.\cite{OConnell:2008,Noroozian:2009,Barends:2010}
One of the simplest ways to mitigate the effects of TLS noise and simultaneously overcome amplifier noise is to drive the resonator with the largest readout power possible.\cite{Gao:2008}  This technique is ultimately limited by the nonlinear response of the resonator.  Potential sources of nonlinearity in thin-film superconducting resonators include a power-dependent
current distribution;\cite{dahm:2002}
quasiparticle production from absorption of readout photons;\cite{visser:114504}
or the nonlinear kinetic inductance
intrinsic to superconductivity.\cite{pippard:210, pippard:547, parmenter:1962}

Virtually all measurements of
microresonator detectors  reported to date have used a readout power below the onset of bifurcation.
Here we demonstrate operation of
a lumped-element microresonator detector
both in the low-power, linear regime and deep in the nonlinear regime
well above the onset of bifurcation.
For most  of our
measurements, the pixel was illuminated with a substantial optical load of $P_{opt} = 6.4$ pW.
For comparison,
at the highest achievable readout power (discussed below) the readout power dissipated in the resonator was $\sim$1.6 pW.  While this is
comparable to
the optical loading, the efficiency for conversion of this power into quasiparticles is expected
and observed\cite{visser:162601}
to be low since the energy of each readout photon is a factor of $\Delta/h f_r = 1.76 k_B T_c/ h f_r = 73$ below the superconducting gap energy.  Much of the
dissipated microwave power may be expected to escape as low-energy, non-pair-breaking phonons,
in which case the quasiparticle population may not change substantially due to the microwave dissipation.
As a result, it is perhaps not entirely surprising that the behavior of our device
even deep into the bifurcation regime is well described by a model that includes
only the nonlinearity of the kinetic inductance.
%

\section{THEORETICAL MODEL AND RESONANCE FITTING}

The
basic principles of superconducting microresonator detector
readout when operating in the linear regime have been extensively described.\cite{Day:2003, Mazin:2004}  The homodyne readout used for this measurement is shown in Fig.\ \ref{figCircuitPixel}(b).
A microresonator
with an intrinsic, unloaded quality factor $Q_i$ and resonance frequency
$\omega_r = 1/\sqrt{LC}$
is coupled to a transmission line,
yielding a coupling quality factor $Q_c$.
A fraction $\alpha$  of the total inductance $L$ is contributed by the kinetic inductance $L_k$
such that $L_k = \alpha L$.  The overall loaded quality factor is given by $Q_r^{-1} = Q_i^{-1} + Q_c^{-1}$.   A signal generator is
used to drive the resonator near its resonance frequency.  The transmitted signal is amplified by a cryogenic amplifier with noise temperature $T_n$ = 6 K, mixed with a copy of the original signal, and digitized.
The resulting complex amplitude of the measured signal is  described by the forward transfer function
\begin{equation}\label{transmissionEqn}
    S_{21} = 1 - \frac{Q_r}{Q_c}\frac{1}{1+2 j Q_r x}
\end{equation}
where
\begin{equation}\label{xStandard}
    x = \frac{\omega_g - \omega_r}{\omega_r}
\end{equation}
is the fractional
detuning
of the readout generator frequency $\omega_g$ relative to the resonance frequency $\omega_r$.  Varying
$x$ by sweeping the generator frequency
traces out a circle in the complex $S_{21}$ plane.
At resonance ($\omega_g = \omega_r, x = 0$),
the circle crosses the real axis at the closest approach to the origin,
while the $S_{21}$ values for all generator frequencies far from resonance fall on the real axis
near unity.

Increasing the readout power results in the onset of nonlinear behavior.  As discussed, the most relevant source of nonlinearity for this device is the nonlinear kinetic inductance
of the superconducting film.
A power-dependent kinetic inductance can be written in terms of the resonator current $I$ with the expression
\begin{equation}\label{nonlinearLk}
    L_{k}(I) = L_{k}(0)[1+I^2/I_*^2 + ...],
\end{equation}
where odd terms are excluded due to symmetry considerations and $I_*$ sets the scaling of the the effect.  $L_{k}(0)$ is the kinetic inductance of the resonator in the low-power, linear limit.

The nonlinear kinetic inductance gives rise to classic soft-spring
Duffing oscillator dynamics.\cite{duffing:1918}
In order to quantitatively account for the power-dependent behavior, it is necessary to replace Eq.\ (\ref{transmissionEqn}) with a transfer function which takes into account the resonance shift $\delta \omega_r$ due to the nonlinear kinetic inductance in Eq.\ (\ref{nonlinearLk}).  The shifted resonance is given by $\omega_r = \omega_{r,0} + \delta \omega_r$ where $\omega_{r,0}$ is the low-power resonance frequency.  Substituting into Eq.\ (\ref{xStandard}), the generator detuning becomes
\begin{equation}\label{xRedefined}
    x = \frac{\omega_g - \omega_{r,0}-\delta \omega_r}{\omega_{r,0}-\delta \omega_r} \approx x_0 - \delta x
\end{equation}
where the approximation is calculated to first order and
\begin{equation}\label{}
    x_0 = \frac{\omega_g - \omega_{r,0}}{\omega_{r,0}}
\end{equation}
is the detuning in the low-power, linear limit.  At a stored resonator energy $E$,
the nonlinear frequency shift $\delta x$ is given by
\begin{equation}\label{deltaX}
    \delta x = \frac{\delta \omega_r}{\omega_{r,0}} = -\frac{1}{2}\frac{\delta L}{L} = - \frac{\alpha}{2}\frac{I^2}{I_*^2} = - \frac{E}{E_*},
\end{equation}
where the scaling energy
$E_* \propto L_k I_*^2 / \alpha^2 $
is expected to be of order the condensation energy of the inductor if $\alpha \approx 1$.

To proceed further,
an expression for
the stored resonator energy at a given readout power and frequency is required.  The
available generator power $P_g$ can be reflected back to the generator,
transmitted past the resonator,
or dissipated in the resonator.  Conservation of
power
can be expressed by
\begin{equation}\label{conservationOfEnergy}
    P_\mathrm{diss} = P_g[1-|S_{11}|^2 - |S_{21}|^2]
\end{equation}
where $P_\mathrm{diss}$ is the power dissipated in the resonator
and $S_{11}$ is the normalized amplitude of the reflected wave.
Noting that $S_{11} = S_{21} - 1$ for a shunt-coupled circuit
and substituting Eq.\ (\ref{transmissionEqn}) into Eq.\ (\ref{conservationOfEnergy}) yields the result
\begin{equation}\label{}
    P_\mathrm{diss} = P_g \left[\frac{2 Q_r^2}{Q_i Q_c} \frac{1}{1+4 Q_r^2 x^2} \right] .
\end{equation}
Using the standard definition of the internal quality factor,
\begin{equation}\label{}
    Q_i = \frac{\omega_r E}{P_\mathrm{diss}}
\end{equation}
the resonator energy is found to be
\begin{equation}\label{resonatorEnergy}
    E = \frac{2 Q_r^2}{Q_c}\frac{1}{1+4 Q_r^2 x^2}\frac{P_g}{\omega_r}.
\end{equation}

\begin{figure}[ht]
   \centering
    \includegraphics[width=82mm]{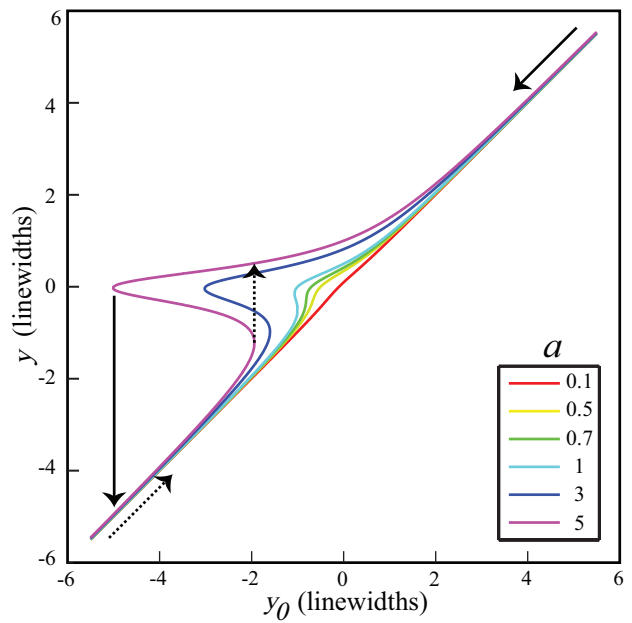}
      \caption{(Color online) Solutions to Eq.\ (\ref{nonlinearEqn}) for a range of the nonlinear parameter $a$.  The solid (dashed) arrows indicate downward (upward) frequency sweeping.  The horizontal scale
      $y_0$
      is the generator detuning measured relative to the low-power resonance frequency $\omega_{r,0}$.  The vertical scale is the the generator detuning $y$ measured relative to the shifted resonance frequency $\omega_r$.  For $a > 4\sqrt{3}/9 \approx 0.8$, $y$ is nonmonotonic in
      $y_0$.
              }
         \label{figYwithYg}
\end{figure}

Eq.\ (\ref{xRedefined}) is an implicit equation for the
power-shifted detuning $x$ as a function of the
generator power $P_g$ and detuning at low power, $x_0$.
To see this, recall from Eq.\ (\ref{xRedefined}) and Eq.\ (\ref{deltaX}) that $x = x_0 + E/E_*$.  Combining this with Eq.\ (\ref{resonatorEnergy}) yields
\begin{equation}\label{implicitX}
    x = x_0 + \frac{2 Q_r^2}{Q_c}\frac{1}{1+4 Q_r^2 x^2}\frac{P_g}{\omega_r E_*} .
\end{equation}
Introducing the variables $y = Q_r x$ and $y_0 = Q_r x_0$ as well as the nonlinearity parameter
\begin{equation}\label{nonlinearParameter}
    a = \frac{2 Q_r^3}{Q_c}\frac{P_g}{\omega_r E_*}
\end{equation}
allows equation~\ref{implicitX} to be rewritten as
\begin{equation}\label{nonlinearEqn}
    y = y_0 + \frac{a}{1+4 y^2}.
\end{equation}
Using the definition of the quality factor $Q_r = \omega_r/\Delta \omega$ where $\Delta \omega$ is the 
linewidth of the resonance, we see that $y = Q_r x = (\omega_g - \omega_r)/\Delta \omega$.  Thus $y$ and $y_0$ are the generator detuning measured in linewidths relative to the power-shifted resonance and the low-power resonance, respectively.  Solutions to Eq.\ (\ref{nonlinearEqn}) for a range of $a$ are shown in Fig.\ \ref{figYwithYg}.  As can be seen from this plot, $y$ becomes nonmonotonic with $y_0$ for $a > 4\sqrt{3}/9 \approx 0.8$.

\begin{figure}[ht]
   \centering
    \includegraphics[width=82mm]{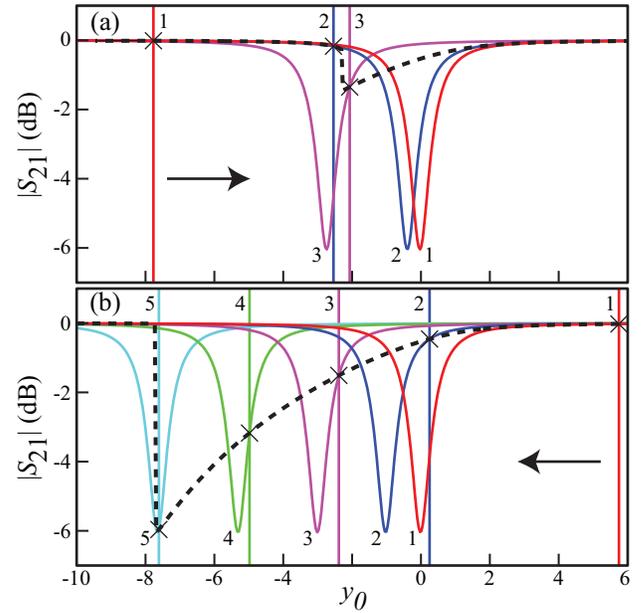}
      \caption{(Color online) Response bifurcation due to feedback.
      Due to the nonlinear kinetic inductance, the resonator current induced by the
      generator shifts the resonance to lower frequency .
      (a)  Upward frequency sweeping.  As the tone enters the resonance (position 2),
      runaway positive feedback causes the resonance to quickly snap to lower frequency (position 3).
      (b) Downward frequency sweeping.  As the generator frequency decreases, negative feedback pushes the resonance toward lower frequencies, away from the generator.
      When the generator goes past the resonance minimum (position 5),
      the resonance snaps back to its unperturbed state (position 1).
      In both subplots, vertical lines indicate various generator frequencies during a
      frequency sweep.
      The intersection of a given measurement tone and the corresponding shifted resonance
      is marked with an X.
      The locus of these intersections
      traces out a hysteretic transfer function (dashed lines) as the frequency is swept upwards
      or downwards.
              }
         \label{calculatedSweeps}
\end{figure}

The origin of the bifurcation is conceptually simple to understand and is visualized in Fig.\ \ref{calculatedSweeps}.  If the generator frequency is swept upwards starting from below the resonance, the resonator current increases as the detuning decreases, and the nonlinear
inductance causes the resonance to shift downward toward the generator frequency, reducing
the detuning further. This process eventually results in a runaway positive feedback condition
as the resonator ``snaps'' into the energized state.
In contrast, sweeping downwards from the high-frequency side results in negative feedback
as the resonance also shifts downwards, away from the generator tone.
The generator tone chases the resonance downward until sweeping past the resonance minimum
when the resonator abruptly snaps back to its non-energized state.
The maximum frequency shift  during downward frequency sweeping will depend on the readout power and reflects the $I^2$ dependence of the kinetic inductance in Eq.\ (\ref{nonlinearLk}).
Notice that smooth downward frequency sweeping allows access to the entire high-frequency side
of the resonance.

\begin{figure}[ht]
   \centering
    \includegraphics[width=82mm]{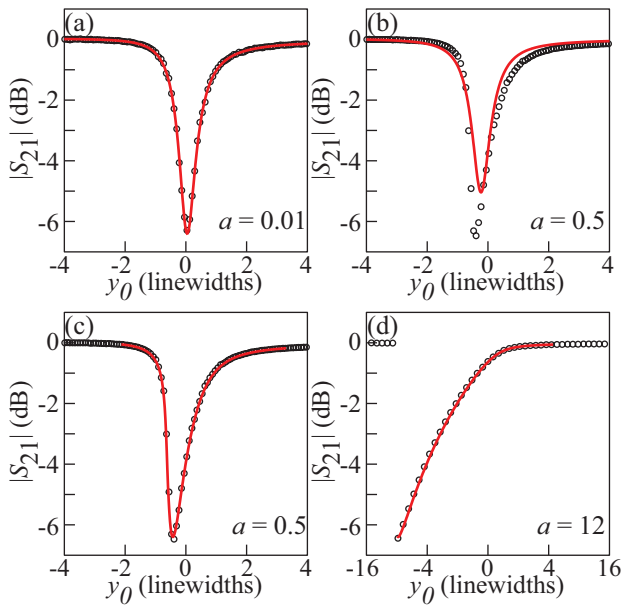}
      \caption{(Color online) Fitting measured resonances for a range of the nonlinear parameter $a$.   In the linear case (a), application of Eq.\ (\ref{transmissionEqn}) yields the desired resonance parameters.  As the readout power is increased (b), direct use of Eq.\ (\ref{transmissionEqn}) is no longer sufficient and results in poor agreement with the data.  Instead, the frequency shifted detuning $x$ can be calculated from Eq.\ (\ref{implicitX}) and substituted into Eq.\ (\ref{transmissionEqn}).  This approach results in good agreement to the data both below (c) and above (d) the onset of bifurcation.
              }
         \label{figFitting}
\end{figure}

Fitting a measured resonance curve yields valuable information including the resonance frequency and quality factors.  A fit to Eq.\ (\ref{transmissionEqn}) of a calibrated resonance in the low-power, linear regime under 6.4 pW of optical loading is shown in Fig.\ \ref{figFitting}(a).  From this fit, we find $Q_i = 8.7 \times 10^5$, $Q_c = 8.1 \times 10^5$ and the low-power resonance frequency is $f_{r,0} = 1.06$ GHz.  A blind application of Eq.\ (\ref{transmissionEqn}) in the nonlinear regime results in a poor fit to the resonance, as exhibited in Fig.\ \ref{figFitting}(b).  Instead, the frequency shifted detuning $x$ at the appropriate $P_g$ must be found from Eq.\ (\ref{implicitX}) and substituted into Eq.\ (\ref{transmissionEqn}).  The
nonlinear energy scale $E_*$
can be determined from Eq.\ (\ref{nonlinearParameter}) by carefully measuring the generator power at the onset of bifurcation ($a \approx .8$).  The results both below and well above bifurcation can be seen in Fig.\ \ref{figFitting}(c)-(d).  Here, the calibration parameters and the low-power fitted quality factors have been fixed.  Only the frequency shifted detuning $x$ has been substituted into Eq.\ (\ref{transmissionEqn}) yielding good agreement with the
measured data
over a broad range of generator powers.

The maximum achievable readout power in this device was limited by the abrupt onset of additional dissipation in the resonator. Switching occurred at 18 dB above the bifurcation power in this device at a dissipated power of $P_\mathrm{diss} >$ 1.6 pW.  In the $S_{21}$ plane, the new state traces out a circle with a smaller diameter than the original resonance circle.  While the source of this additional dissipation is currently under investigation, we can speculate that at a sufficiently high readout photon density in the resonator, multi-photon absorption by the quasiparticles can result in emission of phonons with energy $h\nu>2\Delta$.  These high energy phonons can subsequently break Cooper pairs resulting in an increased quasiparticle density.\cite{goldie:015004}  All measurements presented here were taken below the emergence of this behavior.  Comparing Fig.\ \ref{figFitting}(a) and (d), it is evident that the depth of the transfer function on resonance remained constant from the linear regime to deep within the bifurcation regime.  This indicates that the device dissipation is readout power independent.  Thus, before the emergence of an additional device state, the nonlinear effects can be completely understood as being reactive and not dissipative in nature.

The condensation energy of the inductor is given by $E_\mathrm{cond} = N_0 \Delta^2 V_L/2$ where $N_0$ is the single spin density of states at the
Fermi
energy, $\Delta \approx 3.5 k_B T_c /2$ is the superconducting gap, and $V_L$ is the volume of the inductor.\cite{tinkham:1975}  $E_\mathrm{cond} = 3 \times 10^{-13}$ J for this device.  This is a factor of 5 greater than the
energy scale $E_*$
determined from the onset of bifurcation.
Additional measurements of resonator detectors suggest $E_*$ and $E_\mathrm{cond}$ are comparable for a variety of inductor volumes and critical temperatures.\cite{mckenney:84520, shirokoff:84520}
While $E_\mathrm{cond}$ and $E_*$ are in reasonably good agreement,
caution must be exercised when comparing these quantities,
because knowledge of the absolute power level at the resonator is difficult to ascertain.
This uncertainty arises from
the changing electrical attenuation of the microwave coaxial cable
upon cooling and, in particular, impedance mismatch between the 50 $\Omega$ coaxial transmission line and the on-chip coplanar waveguide.  Additionally, the superconducting gap $\Delta$ is current dependent and deviates from the zero current value near bifurcation.\cite{anthore:127001}

\section{Optical response and noise equivalent power}

\begin{figure}[ht]
   \centering
    \includegraphics[width=82mm]{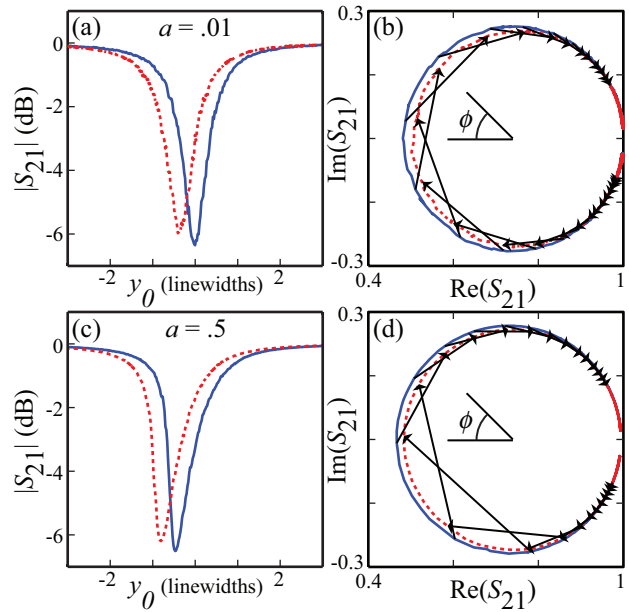}
      \caption{(Color online) Measured LEKID response below bifurcation to increasing the optical illumination from 6.4 pW (solid, blue) to 7 pW (dashed, red). (a) In the linear case ($a$ = .01) the resonance shifts to lower frequencies and becomes shallower.  (b) Response in the $S_{21}$ plane.  The resonance angle $\phi$ shown in this plot is defined such that on resonance $\phi = 0$ and is positive for generator frequencies greater than $\omega_r$ (ie $\phi = 0$ for $x=0$ and $\phi > 0$ for $x>0$).  Arrows indicate the measured displacement at a fixed generator frequency.  In the linear case, the increased optical loading results in a symmetrical clockwise motion about resonance $\phi = 0$.  (c)  As the readout power is increased into the nonlinear regime ($a$ = .5), the resonance is compressed toward lower frequencies.  As explained in the main text, this distortion be understood in terms of reactive feedback.  (d) In the complex $S_{21}$ plane, the feedback causes an augmented response for $\phi <0$ and diminished response for $\phi>0$.
              }
         \label{figS21Signal}
\end{figure}

\begin{figure}[ht]
   \centering
    \includegraphics[width=82mm]{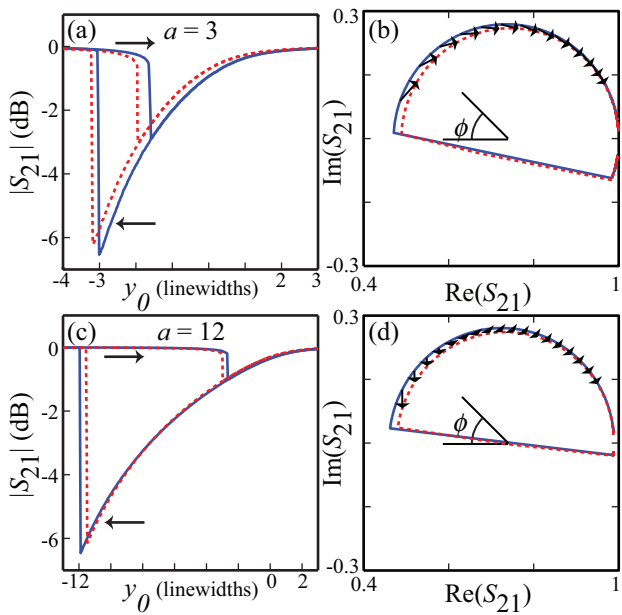}
      \caption{(Color online) Measured LEKID response in the bifurcation regime to a change in optical loading from from 6.4 pW (solid, blue) to 7 pW (dashed, red).  (a)  Above bifurcation ($a$ = 3), feedback results in a reduction in the frequency shift.  As indicated by the arrows, the upper curves were taken while upward sweeping while the lower curves were taken while downward sweeping.  (b) As a result of reactive feedback, the response in $S_{21}$ is considerably diminished but maintains a clockwise rotation.  Here, only the downward frequency sweep is shown.  (c) At sufficient readout power, the reduction in the current due to the increased dissipation $Q_i^{-1}$ causes the resonance to shift to \emph{greater} frequency upon an increase in optical loading.  Notice that at some generator detuning, there is in fact no frequency response.  (d)  The resulting motion in $S_{21}$ in this case reverses sense to a counterclockwise rotation.
              }
         \label{figS21SignalB}
\end{figure}

During detection, incident energy absorbed by the detector breaks Cooper pairs creating quasiparticles.  As predicted by the Mattis-Bardeen theory,\cite{mattis:412} this increases the dissipation $\delta Q_i^{-1}$ and kinetic inductance of the superconducting film.  The resulting behavior in the linear regime can be seen in Fig.\ \ref{figS21Signal}(a)-(b). The increased dissipation and inductance decrease the resonance depth and frequency, respectively. Increasing the readout power results in the onset of nonlinear behavior as shown in Fig.\ \ref{figS21Signal}(c)-(d).  The complex response in $S_{21}$ becomes asymmetric about resonance ($\phi = 0$), diminishing for generator frequencies above the resonance frequency ($x > 0$).  The reduction can be understood in terms of reactive feedback.  Increased optical illumination augments the kinetic inductance, shifting the resonance toward lower frequencies.  The generator tone, set to a fixed frequency, is then situated further out of the shifted resonance reducing the resonator current.  The nonlinear kinetic inductance is consequently reduced causing the resonance to move back to higher frequencies.  This process continues until a stable equilibrium is achieved.  For $x < 0$, the feedback produces the opposite effect resulting in an augmented response.

In order to probe the resonance above bifurcation, the two branches of the response can be accessed experimentally by smoothly sweeping the generator frequency in either the upward or downward sense.  As indicated in Fig.\ \ref{figCircuitPixel}(b), we have accomplished bidirectional frequency sweeping using a voltage controlled oscillator for the signal generator.  Small voltage steps and low-pass filtering ensured smooth frequency sweeping.  The measured transfer function above bifurcation can be seen in Fig.\ \ref{figS21SignalB}(a)-(b).  Due to the runaway positive feedback described above, most of the resonance circle in the complex plane is inaccessible while upward frequency sweeping.  In contrast, nearly the entire upper half of the resonance circle ($\phi > 0$) is accessible during downward frequency sweeping above bifurcation.

Usually in the linear regime, changes in the reactance produce a shift in the resonance frequency but do not affect the resonance depth.  Similarly,  dissipation perturbations only change the resonance depth.  In contrast, in the nonlinear regime dissipation perturbations can produce a frequency response.     As discussed, on the high frequency side of the resonance increased optical loading increases the kinetic inductance while reactive feedback tends to stabilize the resonance against shifting toward lower frequencies.  The additional loading however also increases the dissipation.  The resonance depth and resonator current decrease, thus shifting the resonance toward higher frequencies.  This effect becomes increasingly important as the readout power is increased.  At sufficient resonator currents the \emph{dissipative} frequency response can dominate.  As shown in Fig.\ \ref{figS21SignalB}(c)-(d), the resonance in this case will instead move to higher frequencies with increasing optical loading producing a reversal in chirality in the complex response plane.

In order to determine the expected optical response and noise in the nonlinear regime, we have calculated the first-order perturbation to the power-shifted generator detuning $\delta x$ to changes in the low-power resonance frequency $\omega_{r,0}$ and dissipation $\delta Q_i^{-1}$ using Eq.\ (\ref{implicitX}).  This results in the expression
\begin{equation}\label{perturbation1}
    \delta x = \frac{\delta x_0 + \left(\frac{\partial \tilde{E}}{\partial Q_i^{-1}}\right) \delta Q_i^{-1}}{1 - \frac{\partial \tilde{E}}{\partial x}}
\end{equation}
where $\tilde{E} = E/E_*$.  The derivatives in Eq.\ (\ref{perturbation1}) can be calculated from Eq.\ (\ref{resonatorEnergy}).  The results are
\begin{equation}\label{pEpx}
    \frac{\partial \tilde{E}}{\partial x} =  \left[\frac{1}{1+x}+\frac{-8 Q_r^2 x}{1+4 Q_r^2 x^2}\right]\tilde{E}
\end{equation}
and
\begin{equation}\label{pEpQiInv}
    \frac{\partial \tilde{E}}{\partial Q_i^{-1}} = \frac{- 2 Q_r \tilde{E}}{1+4 Q_r^2 x^2} .
\end{equation}
Note that these results are only valid for slow variations of $\omega_{r,0}$ and $\delta Q_i^{-1}$ well below the adiabatic cutoff frequency $\omega_r/2Q_r$.  At higher frequencies, the ring-down response of the resonator and the feedback must be considered.\cite{Zmuidzinas:2012}  This is not a limitation in practice as current instruments utilizing low-temperature detectors are normally concerned with measurement signals well below the adiabatic cutoff frequency and use low pass filtering to eliminate higher frequency noise.

\begin{figure}[ht]
   \centering
    \includegraphics[width=82mm]{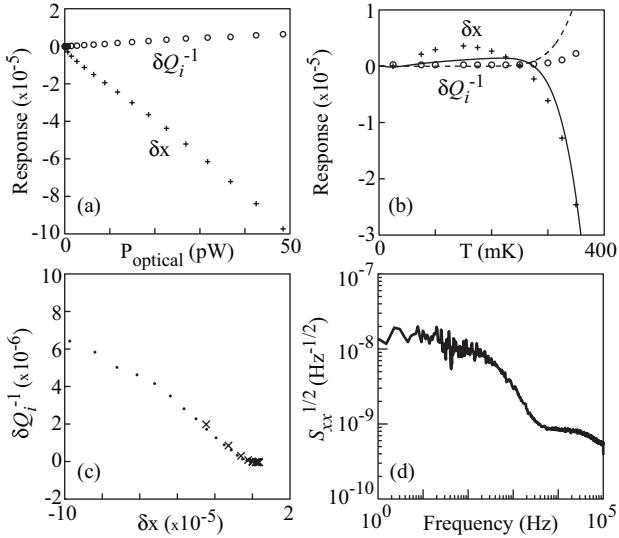}
      \caption{(Color online)  Measured response and noise in the linear regime.  (a) Fractional frequency shift $\delta x$ (+) and dissipation shift $\delta Q_i^{-1}$ (o) as a function of blackbody illumination. (b) Fractional frequency shift $\delta x$ (+) and dissipation shift $\delta Q_i^{-1}$ (o) as a function of mixing chamber temperature.  The initial rise in $\delta x$ at low temperature can be understood from TLS effects.  A fit of $\delta x$ to the Mattis-Bardeen theory\cite{mattis:412} including a TLS contribution\cite{Gao:2008} is shown (solid line) along with the corresponding prediction for $\delta Q_i^{-1}$ (dashed line).  The discrepancy between the measured data and theory has been observed in numerous TiN and NbTiN devices and is an active area of research.\cite{Driessen:2012, gao:142602} (c) Dissipation shift $\delta Q_i^{-1}$ versus fractional frequency shift $\delta x$.  Here, data taken by adjusting the blackbody illumination are marked with a dot (.) and data taken by adjusting the mixing chamber temperature is indicated with an x.  (d) Fractional frequency noise $S_{xx}^{1/2}$ measured near resonance in the linear regime ($\phi \approx 0$, log$(a) = -1.7$).
              }
         \label{figResponse}
\end{figure}

\begin{figure}[ht]
   \centering
    \includegraphics[width=82mm]{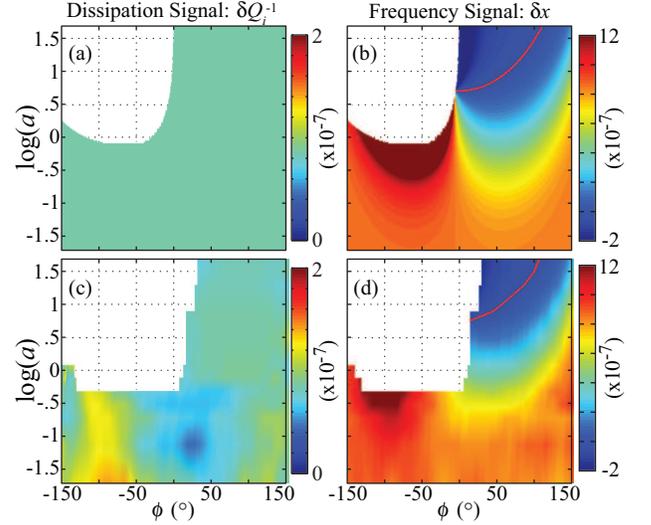}
      \caption{(Color online)  LEKID response to a change in optical loading from 6.4 to 7 pW.  (a) Calculated dissipation response $\delta Q_i^{-1}$. (b) Calculated frequency response $\delta x$ obtained using Eq.\ (\ref{perturbation1}).  (c) Measured dissipation response. (d) Measured frequency response.  The increased optical illumination produces excess quasiparticles.  As can be seen in (a,c), this causes a uniform increase in the resonator dissipation $\delta Q_i^{-1}$ independent of the readout power or generator detuning angle $\phi$.  The increased $n_{qp}$ also produces an additional reactance which causes a low-power frequency shift $\delta x_0$.  At higher powers, the observed frequency response depends on both $\delta Q_i^{-1}$ and $\delta x_0$ as well as a feedback term $(1 - \frac{\partial \tilde{E}}{\partial x})$ according to Eq.\ (\ref{perturbation1}).  Shown in (b,d), above log$(a) > -.2$ is the bifurcation regime where a large portion of the resonance is inaccessible ($\phi < 0$) and the frequency response is suppressed by the feedback.  The red contours indicates $\delta x = 0$.  Above this contour and at small, positive $\phi$, the frequency response contributed by $\delta Q_i^{-1}$ dominates and $\delta x < 0$, indicating the resonance has moved to higher frequencies under the increased illumination.  Note that for the measurement, an automatic data taking procedure utilized a fixed frequency step for $\omega_g$ while downward sweeping.  The steep dependence of $\phi$ on $\omega_g$ near resonance resulted in the region at small positive $\phi$ above bifurcation to not be accessed.   Future measurements can decrease the frequency step for $\omega_g$ when approaching $\omega_r$ to obtain small, fixed steps in $\phi$.
              }
         \label{figSignal}
\end{figure}

\begin{figure}[ht]
   \centering
    \includegraphics[width=82mm]{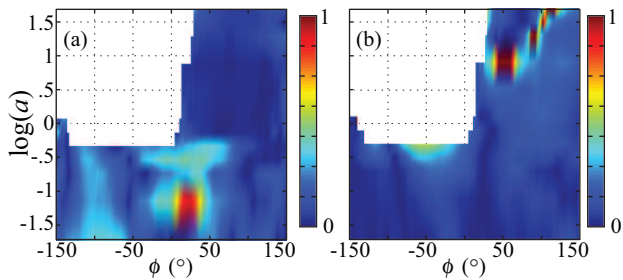}
      \caption{(Color online)  Comparison of calculated and measured response.  The fractional error between the calculated and measured response is computed using the data shown in Fig.\ \ref{figResponse} and the equation $|$measured response - calculated response$|$/measured response.  Shown are the fractional error in (a) the dissipation response $\delta Q_i^{-1}$ and (b) the fractional frequency response $\delta x$.  For $\delta x$, the stripe of large errors in the region log$(a) > .5$ is the result of division by a diminishing measured $\delta x$.  Summing over all the data shown in (a), the root-mean-square (RMS) fractional error for the dissipation response is 0.20.  For (b), the RMS error is 0.22 excluding points where the measured $\delta x$ approaches 0.
              }
         \label{figFractionalError}
\end{figure}

In order to apply Eq.\ (\ref{perturbation1}), appropriate values for $\delta x_0$ and $\delta Q_i^{-1}$ must be provided.  The measured optical response in the low-power linear regime is shown in Fig.\ \ref{figResponse}(a).  A change in optical illumination from $P_{opt} = 6.4$ to 7 pW can be seen to perturb both the resonance frequency and dissipation, with values $\delta x_0 = 8 \times 10^{-7}$ and $\delta Q_i^{-1} = 8 \times 10^{-8}$ respectively.  For comparison, the measured thermal response is shown in Fig.\ \ref{figResponse}(b).  For both the optical and thermal response, $\delta Q_i^{-1}$ is plotted as a function $\delta x_0$ in Fig.\ \ref{figResponse}(c).  The similarity of the two curves indicates that the device response is independent of the source of excess quasiparticles. The relative frequency-to-dissipation response is found to have a ratio $\delta x_0/\delta Q_i^{-1} \approx 10$.  Applying these results to Eq.\ (\ref{perturbation1}), the expected response for our device is given in Fig.\ \ref{figSignal} along with the corresponding measurement result.  The response was obtained both on resonance ($\phi = 0$) and for detuning up to $\phi = \pm 150^{\circ}$.  As previously mentioned, at sufficient resonator currents the dissipative frequency response to increased optical loading can result in the resonance shifting to \emph{higher} frequencies.  The crossover to this behavior is indicated by a red contour where $\delta x = 0$.

\begin{figure}[ht]
   \centering
    \includegraphics[width=82mm]{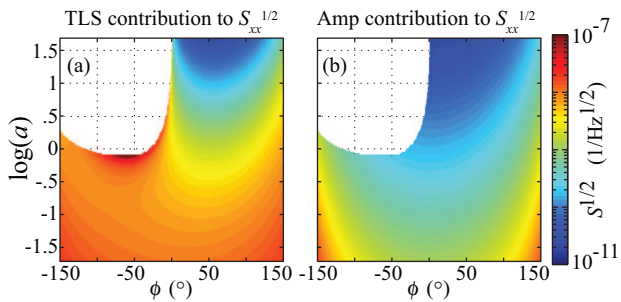}
      \caption{(Color online) Calculated contributions to the measured LEKID fractional frequency noise $S_{xx}^{1/2}$. (a) TLS contribution.  Here we have assumed that the TLS noise is suppressed by frequency feedback but not dissipative feedback in the nonlinear regime.  Additionally, we have assumed the TLS noise fluctuations are suppressed as $P_\mathrm{diss}^{1/4}$ as previously observed by Gao et. al.\cite{Gao:2008a, Gao:2008b} (b) Amplifier contribution assuming a 6 K noise temperature.  The TLS contribution dominates throughout nearly the entire parameter space.
              }
         \label{figFNoiseContributions}
\end{figure}

\begin{figure}[ht]
   \centering
    \includegraphics[width=82mm]{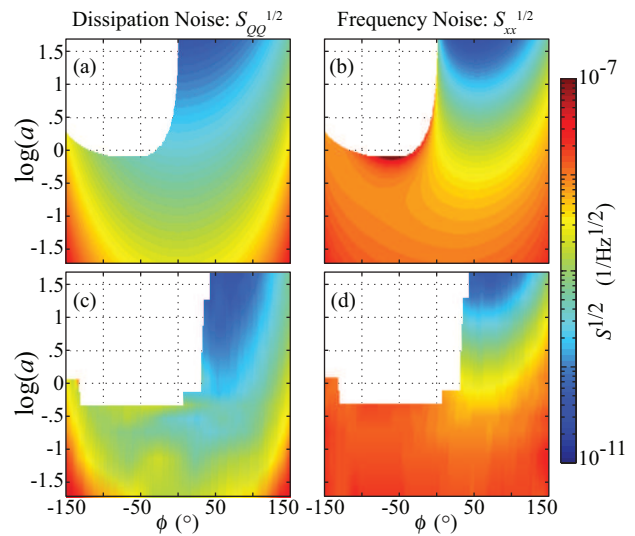}
      \caption{(Color online) Device noise at 10 Hz. (a) Calculated dissipation noise. (b) Calculated frequency noise obtained using Eq.\ (\ref{perturbation1}).  (c) Measured dissipation noise. (d) Measured frequency noise.  For (a,c), the observed dissipation noise improvement with increasing $a$ is due to the straight-forward improvement in signal-to-noise resulting from using an increased generator power $P_g$ relative to the fixed amplifier noise temperature $T_n = 6$ K.  The RMS fractional error comparing (a) and (c) is 0.84.   For (b,d), the improvement in the frequency noise with increasing $a$ results from a combination of an increasing $P_g$ relative to the amplifier noise, a decrease in the TLS noise with stored resonator energy, and the frequency feedback term in the denominator of Eq.\ (\ref{perturbation1}).  The frequency feedback above log$(a) > -.2$ is maximum around $\phi = 45^{\circ}$ and diminishes for smaller $\phi$ resulting in increased $\delta x$ fluctuations near $\phi = 0$.  The RMS fractional error comparing (b) and (d) is 0.75.  In order to determine the measured noise, a time stream of fractional frequency perturbations $\delta x(t)$ and dissipation perturbations $\delta Q_i^{-1}(t)$ were taken at a variety of detuning angles $\phi$ and values of the nonlinearity parameter $a$.  The square root of the measured power-spectral density was then obtained, yielding the frequency noise $\sqrt{S_{xx}}$ and dissipation noise $\sqrt{S_{QQ}}$ respectively.  As noted in the caption of Fig.\ \ref{figSignal}, the use of fixed frequency steps for $\omega_g$ rather than small, fixed steps in $\phi$ while downward sweeping resulted in the measurement not accessing the region at small positive $\phi$ above bifurcation.
              }
         \label{figNoise}
\end{figure}

In order to calculate the expected device noise, both two-level system and amplifier contributions must be considered.  The fractional-frequency noise of the device at low power, given by the square root of the measured power-spectral density $\sqrt{S_{xx}}$ of the fractional frequency noise $\delta x(t)$, is shown in Fig.\ \ref{figResponse}(d).  From this, a value of $\delta x_0 = 1 \times 10^{-8}$ 1/Hz$^{1/2}$ at 10 Hz can be used in Eq.\ (\ref{perturbation1}) to calculate the expected frequency noise in the nonlinear regime.  Additionally, we assume that this value of $\delta x_0$ is suppressed as $P_\mathrm{diss}^{1/4}$ as previously observed by Gao et. al.\cite{Gao:2008a, Gao:2008b}  The TLS fluctuations in the capacitor dielectric which produce this frequency noise have not been observed to produce dissipation fluctuations.\cite{Gao:232508}  Thus for the TLS noise $\delta Q_i^{-1} = 0$ and no dissipative frequency response is possible.  For the amplifier contribution, we have assumed a $T_n =$ 6 K noise temperature of our cryogenic amplifier.  The fluctuations in $S_{21} = (4 k T_n/P_g)^{1/2}$ are then converted to dissipation and frequency fluctuations.  Both the calculated TLS and amplifier frequency noise contributions are shown in Fig.\ \ref{figFNoiseContributions}.  These can be summed in quadrature to yield the total device frequency noise.  Both the calculated dissipation and total frequency noise are shown alongside the measured results in Fig.\ \ref{figNoise}.

\begin{figure}[ht]
   \centering
    \includegraphics[width=82mm]{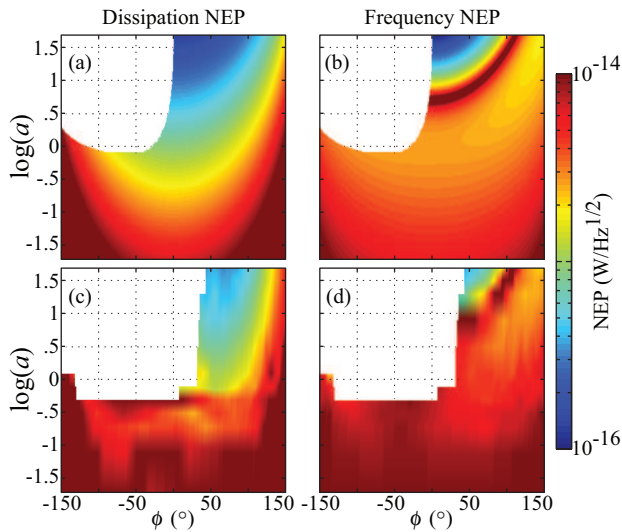}
      \caption{(Color online) Calculated and measured NEP at 10 Hz under 6.4 pW of optical loading. (a) Calculated dissipation NEP.  (b) Calculated frequency NEP.  (c) Measured dissipation NEP.  (d) Measured frequency NEP.  Note in both frequency NEP subplots there is a band in the bifurcation region (log$(a) > -.2)$ which exhibits a  dramatically increasing NEP.  As shown in Fig.\ \ref{figSignal}, in this region there is a vanishing frequency response ($\delta x = 0$) resulting in diminished device performance.  In contrast, at high powers and near resonance, the dissipative frequency response dominates.  As the TLS noise has no dissipative contribution, it is simply suppressed by frequency feedback.  This results in a region with a significantly improved NEP$_{freq}$.
              }
         \label{figNEP}
\end{figure}

\begin{figure}[ht]
   \centering
    \includegraphics[width=82mm]{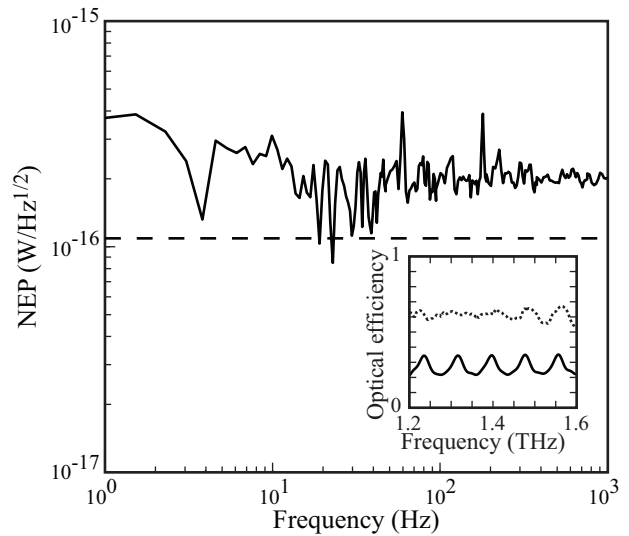}
      \caption{(Color online) Best achieved device noise equivalent power calculated using NEP$^{-2}$ = NEP$_{diss}^{-2}$ + NEP$_{freq}^{-2}$.  This NEP was measured deep in the bifurcation regime at log($a$) = 1.7 and on the high frequency side of the resonance at $\phi = 40^{\circ}$.  Close inspection of Fig.\ \ref{figFNoiseContributions} reveals that under these conditions, the TLS noise contribution to the frequency noise is suppressed below the amplifier contribution.  Thus the NEP shown here is limited by uncorrelated amplifier noise in both quadratures.  The dashed line indicates the expected photon-noise limited NEP$_{photon} = \sqrt{2 P_0 h \nu (1+n_0)}$, where $P_0$ is the optical illumination and $n_0$ is the occupation number.  \textit{Inset} Simulated optical absorption with the current measurement setup (solid, $\eta_{opt} \sim .3$) and after optimization (dashed, $\eta_{opt} \sim .6$).
              }
         \label{figBestNEP}
\end{figure}

Combining the response in Fig.\ \ref{figSignal} with the noise in Fig.\ \ref{figNoise} yields the device noise-equivalent power (NEP) in the dissipation and frequency quadratures shown in Fig.\ \ref{figNEP}.  These independent quadratures may be combined according to NEP$^{-2}$ = NEP$_{diss}^{-2}$ + NEP$_{freq}^{-2}$.  The best NEP of $2 \times 10^{-16}$ W/Hz$^{1/2}$ at 10 Hz was obtained at the highest readout power at a detuning angle of $\phi = 40^{\circ}$ and is shown in Fig.\ \ref{figBestNEP}.  Note that this is a $\sim$10 fold improvement over the best NEP below bifurcation.  We emphasize that this gain is the result of two mechanisms.  First, as previously noted, increasing the readout power decreases the effects of TLS noise while also overcoming amplifier noise.  This straightforward increase in signal-to-noise substantially explains the NEP improvement.  However, this is not the whole story.  Eq.\ (\ref{perturbation1}) provides an additional mechanism for improving NEP$_{freq}$.  Due to the dissipative $\delta Q_i^{-1}$ term in this equation, changes in the quasiparticle density from the optical signal result in both a reactive \emph{and} dissipative frequency response.  At high powers and near resonance, the dissipative frequency response dominates.  However, as the TLS noise has no dissipative contribution, it is simply suppressed by the frequency feedback term $(1 - \frac{\partial \tilde{E}}{\partial x})$.  The difference in the behavior of the frequency response and noise above the onset of bifurcation produces a region at high powers and near resonance with a substantially improved NEP$_{freq}$.

The best measured NEP is a factor of two above photon-noise limited performance for the current optical illumination.  In order to achieve photon-noise limited operation, a number of optimizations can be made.  First, as indicated in the inset of Fig.\ \ref{figBestNEP}, the simulated dual-polarization optical efficiency of this device under the experimental conditions was $\eta_{opt} \approx .3$.  By including an anti-reflection coating, backshort, and tuning the TiN sheet resistance we find that the optical efficiency can be improved to $\eta_{opt} \approx .6$.  Implementing these changes would then give a modest $\times$1.4 improvement in the NEP.  Also, the fractional frequency noise $S_{xx}^{1/2}$ has been observed to decrease linearly with increased temperature.  Thus operating at modestly increased temperatures, while taking care that the thermally generated quasiparticles remain negligible compared with those that are optically generated under expected loading conditions, would provide an improved NEP.  Implementing these changes would potentially allow the current device to operate with photon-noise limited performance under the typical illumination conditions found in ground-based, far-infrared astronomy.

\section{CONCLUSION}

We have characterized the behavior of a lumped-element kinetic inductance detector optimized for the detection of far-infrared radiation in the linear and nonlinear regime.  The device was fabricated from titanium nitride, a promising material due to its tunable $T_c$, high intrinsic quality factor, and large normal state resistivity.  The measurements were performed under 6.4 pW of loading which is comparable to or somewhat less than the expected loading for ground based astronomical observations.  The device was driven nonlinear by a large readout power which is desirable due to the suppression of two-level system noise in the capacitor of the device at high power and the diminishing importance of amplifier noise at large signal powers.  At sufficient readout powers, the transfer function of the detector bifurcates.  By smoothly downward frequency sweeping a voltage controlled oscillator we were able to access the upper frequency side of the resonance.  The best noise equivalent power in this regime was of $2 \times 10^{-16}$ W/Hz$^{1/2}$ at 10 Hz, a $\sim$10 fold improvement over the sensitivity below bifurcation.

Two practical conclusions can be drawn from this work.  First, the onset of bifurcation can be increased simply by decreasing $Q_r$.  This allows operation at high readout powers without necessitating the use of a smooth, downward frequency sweep.  While this provides a mechanism for achieving improved device performance, the decreased $Q_r$ results in each pixel occupying a larger portion of frequency space.  This proportionally decreases the multiplexing factor of the array resulting in increased electronics costs.  Secondly, if the TLS noise of the device can be engineered below the amplifier noise contribution, increased pixel performance can be achieved by operating just below bifurcation on the low-frequency side of the resonance.  As previously observed\cite{monfardini:24} and shown here, the optical frequency response is enhanced in this region.  Meanwhile the amplifier noise is unaffected by the resonator nonlinearity.  The NEP$_{freq}$ is consequently improved.

The results presented are of general interest to the low-temperature detector community focusing on microresonator detectors.  First, the included nonlinear resonator fitting model allows extraction of the useful resonator parameters at large readout powers when the kinetic inductance is the dominant device nonlinearity.  Next, while increasing the readout complexity, the technique of smooth downward frequency sweeping can significantly increase the detector performance compared to operation below the onset of bifurcation while maintaining a high resonator $Q_r$ necessary for achieving dense frequency multiplexing.  Note that this technique can be simultaneously applied to all resonator in an imaging array, shifting the resonances uniformly and preserving the pixel frequency spacing.  Finally, the observation that the scaling energy $E_*$ is of order the inductor condensation energy allows a useful estimate of the onset of nonlinear behavior and hysteresis.  We expect that a variety of experiments, particularly kinetic-inductance based detectors for sub-mm astronomy and dark-matter detection, will benefit from this work.

\begin{acknowledgments}
      The authors wish to thank Teun Klapwijk and David Moore for useful discussions relating to this work.  This work was supported in part by the Keck Institute for Space Science, the Gordon and Betty Moore Foundation.  Part of this research was carried out at the Jet Propulsion Laboratory (JPL), California Institute of Technology, under a contract with the National Aeronautics and Space Administration.  The devices used in this work were fabricated at the JPL Microdevices Laboratory.  L. Swenson acknowledges support from the NASA Postdoctoral Program.  L. Swenson and C. McKenney acknowledge funding from the Keck Institute for Space Science.  \copyright 2012. All rights reserved.
\end{acknowledgments}


\end{document}